\documentclass[10pt]{iopart}

\usepackage[colorlinks=true, linkcolor=blue, citecolor=blue, urlcolor=blue]{hyperref}
\usepackage{graphicx}
\usepackage{cite}

\begin{document}

\title[Topological phase transition in OsCl$_3$ monolayer]{In-plane magnetization orientation driven topological phase transition in OsCl$_3$ monolayer}

\author{Ritwik Das, Subhadeep Bandyopadhyay\footnote{Present address: Theoretical Materials Physics, CESAM,
University of Liège, Liège, Belgium} and Indra Dasgupta}

\address{School of Physical Sciences, Indian Association for the Cultivation of Science,\\
2A \&\ 2B Raja S. C. Mullick Road, Jadavpur, Kolkata 700 032, India}
\ead{sspid@iacs.res.in}
\vspace{10pt}
\begin{indented}
\item[]January 2024
\end{indented}
\begin{abstract}
The quantum anomalous Hall effect resulting from the in-plane magnetization in the OsCl$_3$ monolayer is shown to exhibit different electronic topological phases determined by the crystal symmetries and magnetism. In this Chern insulator, the Os-atoms form a two dimensional planar honeycomb structure with an easy-plane ferromagnetic configuration and the required non-adiabatic paths to tune the topology of electronic structure exist for specific magnetic orientations based on mirror symmetries of the system. Using density functional theory (DFT) calculations, these tunable phases are identified by changing the orientation of the magnetic moments. We argue that in contrast to the buckled system, here the Cl-ligands bring non-trivial topology into the system by breaking the in-plane mirror symmetry. The interplay between the magnetic anisotropy and electronic band-topology changes the Chern number and hence the topological phases. Our DFT study is corroborated with comprehensive analysis of relevant symmetries as well as a detailed explanation of topological phase transitions using a generic tight binding model.
\end{abstract}
\vspace{2pc}
\ioptwocol
\sloppy
\section{Introduction}

Two dimensional (2D) electronic systems under a very strong external out-of-plane magnetic field show the quantum Hall effect (QHE) \cite{QHE1,QHE2,QHE3} due to broken time reversal symmetry (TRS) following the quantization of Landau levels. Some magnetic insulators also exhibit the quantized transverse conductivity where the intrinsic magnetization of the system replicates the external magnetic field, which falls under the category of quantum anomalous Hall systems \cite{RMP_Berry, RMP_AHE1,RMP_AHE2}. Broken TRS is a necessary condition to achieve the Hall effect and the 2D honeycomb lattice-geometry can provide exotic electronic topological phases as for the first time was theoretically demonstrated by F.M. Haldane \cite{Haldane}. In case of TRS breaking insulators, integer-valued localized conducting edge-states appear at the edges of these systems corresponding to the non-trivial topology which can be numerically captured by computing the Berry-curvature properties of electronic band structure \cite{RMP_Berry,BBC1,BBC2}. These dissipationless and topologically protected chiral edge-states have several potential applications in the field of low power electronics and spintronics \cite{Application1,Application2,Application3}.

There are several theoretical predictions for material realization of quantum anomalous Hall effect (QAHE) with out-of-plane magnetization \cite{OPQAHE1,OPQAHE2,OPQAHE3,OPQAHE4}. However recent studies in 2D systems with buckled honeycomb structure show that the in-plane magnetization may also contribute to the QAHE where the mirror symmetries play an important role for the realization of non-trivial topological phases. In these cases, buckling and in-plane magnetization disrupt the in-plane mirror symmetry, while buckling introducing the non-trivial topology in the system. The Onsager relation \cite{AHAFM,Landau} implies that the reversal of the magnetization direction results in a sign-change in the Hall-conductivity. However, the orientation of magnetic moments in these systems control their topological phases. In particular, preserving out-of-plane mirror symmetry results in semimetallic, topological phase transition points and breaking this symmetry leads to the formation of Chern insulators with distinct Chern numbers \cite{IPQAHE1,IPQAHE2,IPQAHE3,IPQAHE4,IPQAHE5}.

Monolayer (ML) OsCl$_3$, on the other hand, is a theoretically predicted in-plane ferromagnetic Chern insulator with the topologically non-trivial ground state of Chern number C = -1 when the Hubbard-U correction of Os-atom is smaller than a critical value of $U_c=$ 0.4 eV \cite{Oc}. However, this critical value depends on lattice parameters and the transition metal ion of the system. Unlike the previous studies of in-plane ferromagnetic QAHE with buckled honeycomb structures, Os-atoms form a flat honeycomb lattice in ML OsCl$_3$, where the Cl-ligands break the in-plane mirror symmetry.

The focus of this study is on the fact whether the ligand environment could replicate the buckling effects on electronic topology. In this paper, we show that the system exhibits topological phase transitions following the Onsagar relation, depending on the existing out-of-plane mirror symmetries. The symmetry driven interplay between the easy-plane magnetic anisotropy and topological phases are systemetically studied where the sign of the Chern number can be tuned by changing the orientation of in-plane magnetic moment. A recent study on NiAsO$_3$ and PdSbO$_3$ \cite{Recent} has also shown that the tunable electronic topology can be manipulated in the planar honeycomb structure by changing the direction of the magnetic moment.

We have employed first-principles calculations to compute the intrinsic contribution of anomalous Hall effect (AHE), and Wannier-based tight-binding (TB) models are used to analyze the topological phases. Further we have shown that a generic TB model can capture the topological phase transition. The paper is organized as follows: In Section \ref{II} the computational details and crystal structure of ML OsCl$_3$ are presented. Section \ref{III} includes the ground state electronic and magnetic properties, tunable topological phases, a generic TB model for the system, symmetry analysis and dependence of electronic and magnetic properties on the orientation of magnetic moment. Finally in section \ref{IV} we have drawn the conclusion.
\section{Computational details and crystal structure}
\label{II}

To study the electronic structure and magnetic properties of ML OsCl$_3$, first-principles DFT calculations with plane-wave based projector augmented wave (PAW) method \cite{PAW} is performed using Vienna ab initio simulation package (VASP) \cite{VASP}. Generalized gradient approximation (GGA) of Perdew-Burke-Ernzerhof \cite{GGA} is considered to treat the exchange and correlation effects. Brillouin zone (BZ) integration is computed using a $\Gamma$-centered $15 \times 15 \times 1$ k-point mesh and a cut-off of 600 eV is used for kinetic-energy of the plane-wave basis. The convergence criterion for self-consistent process of energy minimization is fixed to $10^{-8}$ eV and constrained-moment calculations are done to fix the magnetic moment of Os at a particular direction as implemented in VASP. Wannier90 code \cite{W90_1,W90_2} is used to obtain the TB model based on maximally localized Wannier functions (MLWFs) from VASP output which helps to understand the topological properties of the system. Detail analysis of topological phases and edge states are done following the iterative Greens function approach as implemented in the WannierTools \cite{WT} software.

\begin{figure}[t]
\centering
\includegraphics[scale = 1.0]{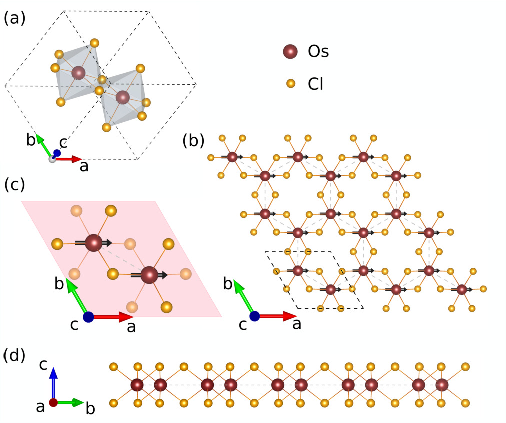}
\caption{The crystal structure of ML OsCl$_3$. \textbf{(a)} The unit cell consists of two-formula units where Cl-ligands form an edge-sharing octahedra. \textbf{(b)} The planar honeycomb lattice formed by Os-atoms is shown with the ground state ferromagnetic configuration. The moments are denoted by black arrows. \textbf{(c)} and \textbf{(d)} Illustration of in-plane mirror symmetry disruption by Cl-ligands relative to the $\hat{M}_z$-mirror, depicted by a pink plane in (c) and a dashed line in (d).}
                \label{fig:structure}
\end{figure}

The crystal structure of ML OsCl$_3$ is shown in figure \ref{fig:structure}(a) and \ref{fig:structure}(b). The nearest neighbour Os-Os bonds form a flat honeycomb structure on the 2D-plane parallel to the primitive lattice translation vectors $\vec{a}$ and $\vec{b}$ where $\mid\vec{a}\mid = \mid\vec{b}\mid = 5.990${\AA}. The Cl-ligands form an edge-sharing octahedra, which is tilted with respect to the crystallographic $\vec{c}$-axis by a finite angle of approximately 55.67$^\circ$ and due to this tilting, the in-plane mirror symmetry is broken with respect to the $\hat{M}_z$-mirror as shown in figure \ref{fig:structure}(c) and (d). Moreover this tilting brings non-trivial topology in the system with in-plane magnetization.

\section{Results and discussions}
\label{III}
\subsection{Ground state electronic structure and magnetic properties}

The primitive unit cell of ML OsCl$_3$ consists of two-formula units (figure \ref{fig:structure}(a)) where the Os-atoms adopt a $d^5$ ($Os^{3+}Cl_3^{-}$) nominal electronic configuration. As a consequence all the states derived from Cl atoms are filled and lie below the Fermi-level ($E_F$). Due to the octahedral environment provided by Cl-ligands, the $d$-levels of the Os-atoms are split into three-fold degenerate $t_{2g}$ and two-fold degenerate $e_g$ states with the crystal-field (CF) splitting of 1.96 eV. The five $d$-electrons of the Os-atom are accommodated in the $t_{2g}$ state in a non-spin-polarized calculation leading to a metallic solution.

\begin{figure*}[htbp]
\centering
\includegraphics[scale = 1]{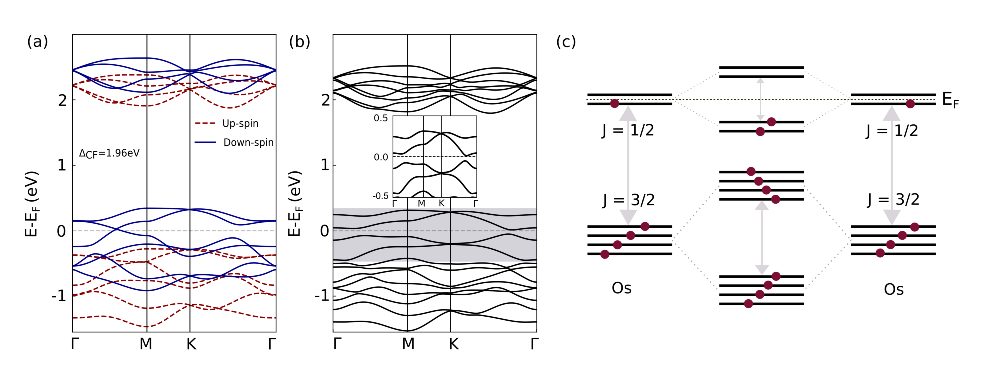}
\caption{Electronic band structure of ML OsCl$_3$. \textbf{(a)} The spin-polarized band structure without SOC. \textbf{(b)} The electronic band structure with SOC. The $J_{eff}$=1/2 sector is shown in the inset. \textbf{(c)} Explanation for the formation of an insulator in presence of SOC.}
                \label{fig:SOC}
\end{figure*}

Collinear spin polarized DFT calculation removes the spin degeneracy in the electronic structure, resulting in a completely filled $t_{2g}$ up-spin channel and partially filled down-spin channel with exchange splitting of 0.55 eV between the two different spin sectors of Os-$t_{2g}$ bands and a total magnetic moment of $1.0\mu_B$/Os, which is consistent with the atomic picture. In this case, the electronic band structure shows half-metallic property with insulating behavior for the up-spin bands and metallic behavior for the down-spin bands, with the band touching at a point between the $\Gamma$ to M high-symmetry paths near $E_F$ as shown in figure \ref{fig:SOC}(a). As there are six $\Gamma-M$ paths in the hexagonal BZ for our system, the spin-polarized spin-down band structure for a pair of Os-atoms in the unit cell has six Weyl points in the 2D BZ which lead to the metallic behavior.

\begin{figure*}[htbp]
\centering
\includegraphics[scale=1.0]{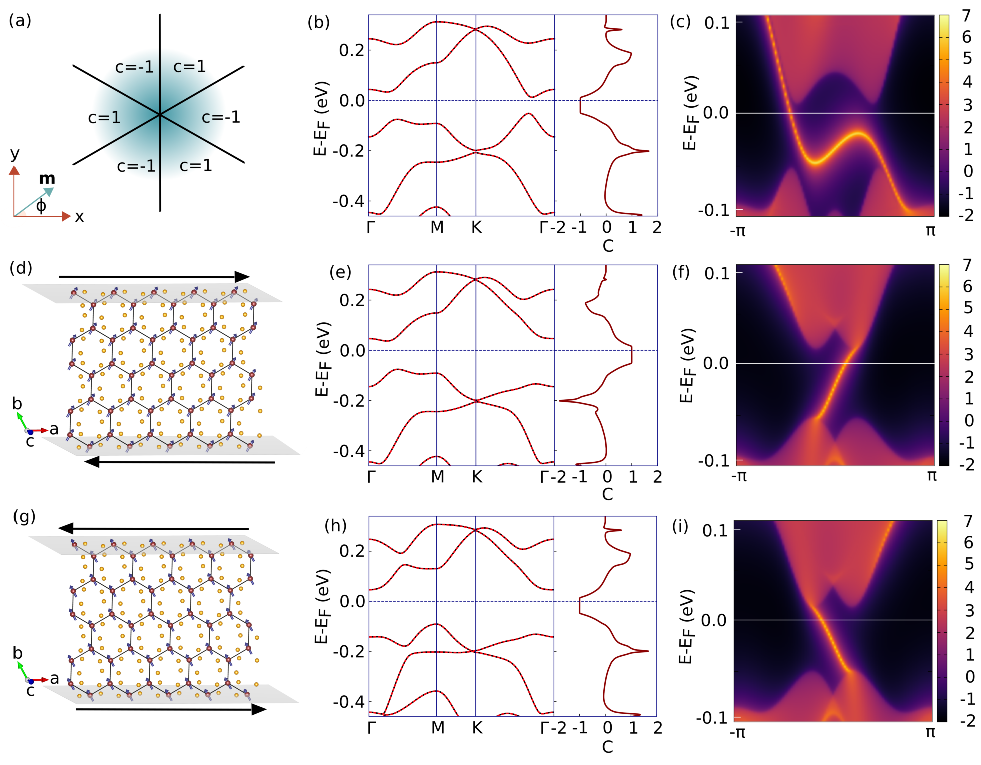}
\caption{The topological properties due to the in-plane variation of magnetic moment of Os-atoms. \textbf{(a)} The in-plane orientation of moment is denoted by the azimuthal angle $\phi$ (inset figure). The Chern number (C) varies between C=-1 and C=+1 depending on $\phi$ with a period of $60^{\circ}$. \textbf{(b)} The DFT-band (solid black lines) and Wannier interpolated bands (dashed red lines) near the $E_F$ and the variation of C with respect to the $E_F$ is shown for ground state ferromagnetic calculation ($\phi$ = 0). \textbf{(c)} The local-DOS for the semi-infinite zigzag boundary condition is ploted for the top-boundary. The cutting plane is parallel to both of the crystallographic axes $\vec{a}$ and $\vec{c}$. \textbf{(d)} The direction of the edge current for $\phi$=$60^{\circ}$ corresponding to C=+1. \textbf{(e)} and \textbf{(f)} As in (b) and (c), but for $\phi$ = $60^{\circ}$. \textbf{(g)}  The direction of the edge current for $\phi$=$120^{\circ}$ corresponding to C=-1. \textbf{(h)} and \textbf{(i)} Analogous to (b) and (c), but for $\phi$ = $120^{\circ}$. The alteration in the slope of the edge states indicates a sign change in C obeying the bulk-boundary correspondence.}
\label{fig:topology}
\end{figure*}

In the presence of spin orbit coupling (SOC), the system becomes insulating, and DFT calculations show that the ground state of the system is a in-plane noncollinear ferromagnet, with the magnetic moments tilted along the crystallographic $\vec{a}$-direction. The electronic band-structure for the ground state in the presence of SOC is shown in figure \ref{fig:SOC}(b). As the atomic SOC of Os is high, it splits the $t_{2g}$ levels further into $J_{eff}$=3/2 and $J_{eff}$=1/2 levels. Assuming two Os-atoms in the unit cell as dimer, the bonding and antibonding splitting of $J_{eff}$=1/2 levels make the system an insulator even in the absence of Hubbard U as schematically shown in figure \ref{fig:SOC}(c). The CF splitting between $t_{2g}$ and $e_g$ levels is much larger than the band-gap $\Delta \approx 73$meV of the system arrising between $J_{eff}$=1/2 bands (figure \ref{fig:SOC}(a) inset). All the DFT calculations are done with an effective value of Hubbard-U=0 because beyond a critical value $U_c$, the system becomes a topologically trivial Mott insulator, as shown in \cite{Oc}.
\subsection{Tunable electronic topology}

To investigate the topological properties of the system, a multiband TB model is computationally constructed using Wannier orbitals that are based on all the d orbitals of the Os-atoms. It has been verified that including the s and d orbitals of Os, as well as the s and p orbitals of Cl-atoms, in the calculations yields similar results, as only the d-states close to the Fermi level are responsible for inducing the non-trivial topology in the system.

\begin{figure}[t]
\centering
\includegraphics[scale=1.0]{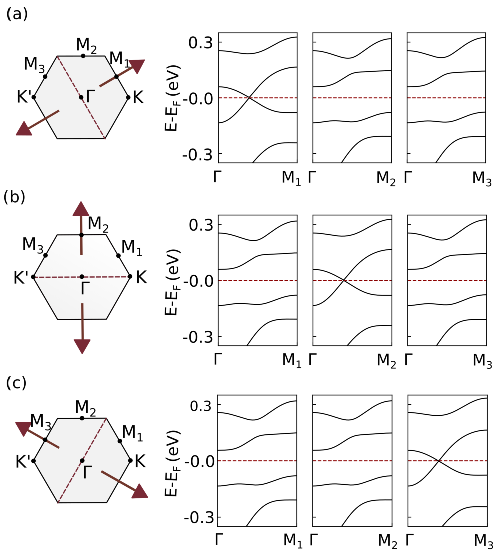}
\caption{Band structures of the ML OsCl$_3$ at the topological phase transition points. \textbf{(a)} The left most side figure shows the moment orientation in the first BZ for $\phi=30^{\circ}$,$210^{\circ}$ and right side figures show the electronic band structures for different high-symmetry lines. The corresponding symmetry preserving out-of-plane mirrors are denoted by dotted lines in the BZ \textbf{(b)} and \textbf{(c)} Same as (a), but for $\phi = 90^{\circ}$,$270^{\circ}$ and $150^{\circ}$,$330^{\circ}$ respectively.}
\label{fig:topology_02}
\end{figure}

The change in topological phases via changing the direction of magnetic moments attached to Os-atoms are presented in figure \ref{fig:topology}. Over a certain range of moment orientation, the Chern number C remains constant and topological phase transition happens for some particular values of $\phi$ (see figure \ref{fig:topology}(a)), where $\phi$ denotes the azimuthal angle made by the moment $\vec{m}$ with respect to the ground state moment orientation along crystallographic $\vec{a}$-axis (same as the x-direction in figure \ref{fig:topology}(a)). The polar angle is always kept at $\theta=\pi/2$ as we are considering only the in-plane variation of moment. The non-zero value of C throughout the band-gap ($C=-1$) for $\phi=0$ as shown in figure \ref{fig:topology}(b) and the one conducting edge state reflected in the local-DOS for finite system in figure \ref{fig:topology}(c) reveals the non-trivial topology of the ground state and the bulk-boundary correspondence. It turns out that the C changes sign for every $60^{\circ}$ change in $\phi$ starting from the ground state at $\phi=0$. The direction of edge current will change accordingly. We have briefly shown these behavior for two different Chern insulating phases, namely for $\phi=60^{\circ}$ and $\phi=120^{\circ}$ in figure \ref{fig:topology}(d)-\ref{fig:topology}(f) and \ref{fig:topology}(g)-\ref{fig:topology}(i) respectively. 

In ML OsCl$_3$, the above mentioned topological phase transition happens due to the presence of easy-plane magnetic anisotropy in the system which arises from the strong SOC of Os. All the topological properties are hosted by the $J_{eff}$=1/2 bands which are situated near $E_F$ (figure \ref{fig:SOC}). To change the topology of the system, the tuning parameter of the Hamiltonian should traverse a non-adiabatically connected path. In this case, the direction of the in-plane localized moments of Os-atoms act as the tuning parameter and whenever there exists a mirror symmetry in the system, the conduction and the valence band touch each other. When the band-gap disappears for these particular orientation of moments, the Berry phase becomes ill-defined \cite{IBC}, indicating the non-adiabatic phases of the system. These topological phase transition points occur at angles of $\phi = 30^{\circ}$, $90^{\circ}$, $150^{\circ}$, $210^{\circ}$, $270^{\circ}$ and $330^{\circ}$. The electronic band structures for these angles are illustrated in figure \ref{fig:topology_02}. It should be noted that for these specific orientations, the system maintains out-of-plane mirror symmetry. The cross-sections of these out-of-plane mirror planes, parallel to the crystallographic $\vec{c}$-axis and perpendicular to the honeycomb plane, are depicted as black dotted lines within the Brillouin Zone (BZ) in Figure \ref{fig:topology_02} which clearly demonstrates that the in-plane magnetic moments are perpendicular to the out-of-plane mirrors for the non-adiabaic phases. Depending on the orientation of the moments and the corresponding out-of-plane mirror symmetries, the band-gap closes at different high-symmetry paths as explained below.

At the non-adiabatic phases, the system transforms to a semi-metallic state. In these states, the conduction and valence bands touch at two distinct points. These points are located along the high-symmetry paths between the $\Gamma$ and $M$ points in the BZ. There are six high symmetry points $M$ in the first BZ of honeycomb structure, among which, three are depicted in figure \ref{fig:topology_02} as $M_1$,$M_2$ and $M_3$. Other three $M$-points can be obtained by taking inverse vectors of $\overrightarrow{\Gamma M_i}$ (i=1,2 or 3) as they are connected through inversion operation. In absence of SOC, all the electronic properties along these six $\Gamma-M$ paths are same, whereas in presence of SOC, the magnetic anisotropy makes these M-paths nonequivalent. For instance, with magnetic moments oriented along the $\overrightarrow{\Gamma M_1}$ direction, a two-fold degenerate point (Weyl-point in 2D) \cite{WHSM_1,WHSM_2,HDP} emerges on the $\Gamma-M_1$ path and its inversion counterpart appears along the line inverse to $\overrightarrow{\Gamma M_1}$ with respect to $\Gamma$-point. The other four high-symmetric M-points remain equivalent with finite band-gap as shown in figure \ref{fig:topology_02}(a). The scenario is analogous for moment orientations along $\overrightarrow{\Gamma M_2}$ and $\overrightarrow{\Gamma M_3}$ owing to the retention of out-of-plane mirror symmetry (figure \ref{fig:topology_02}(b) and (c)). From this result, it is clear that topological phase transition happens in this system when the moments are oriented along these $\Gamma-M$ high-symmetry lines. One degenerate point for each case is shown in figure \ref{fig:topology_02} and the other can be obtained using inversion operation \cite{HDP}.

\begin{figure}[t]
\centering
\includegraphics[scale = 1]{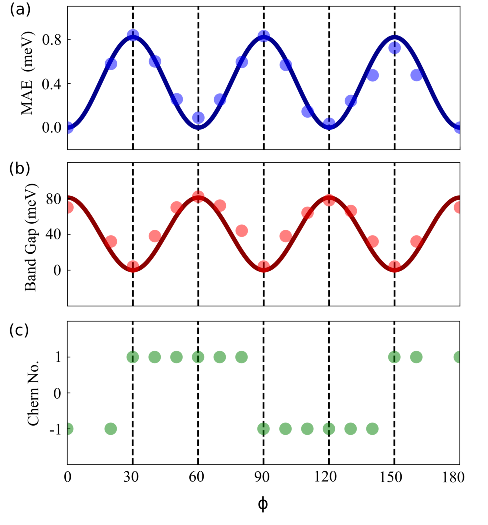}
\caption{\textbf{(a)} The relation between easy-plane magnetic anisotropy and moment orientation $\phi$. \textbf{(b)} Band-gap variation with $\phi$. \textbf{(c)} The change in Chern number C with $\phi$.}
                \label{fig:mae}
\end{figure}

The easy-plane magnetic anisotropy energy (MAE) per unit cell of the ML OsCl$_3$ is the energy difference between $E(\phi)$ and $E(\phi=0)$. Figure \ref{fig:mae}(a) and (b) shows the variation of MAE and the band-gap in ML OsCl$_3$ as a function of in-plane moment orientation ($\phi$), revealing a correlation where larger MAE corresponds to a smaller band-gap. The MAE follows an $A sin^2(3\phi)$ curve with an amplitude $A=0.85 meV$ and a period of $60^{\circ}$, mirroring the oscillation period of the Chern number C. Conversely, the band-gap oscillates in an exactly opposite manner, exhibiting a larger amplitude $A^\prime=81.5 meV$. Changes in C corresponding to these variations are presented in figure \ref{fig:mae}(c). At transition points, maximum MAE coincides with vanishing band-gaps. Since larger band-gaps are energetically favoured, this suggests a preference for broken out-of-plane mirror symmetry in the ground state as evident from figure \ref{fig:mae}(a) and (b).
\subsection{Generic tight-binding Hamiltonian and comparison with the DFT results}

In this subsection, we consider a minimal TB-Hamiltonian which was considered for buckled honeycomb systems to explain the electronic topology for in-plane ferromagnets\cite{IPQAHE3,Main_tb}. 
\vspace{0.2em}
\begin{samepage}
\begin{eqnarray}
\label{TB}
\hat{H} = && -t\sum\limits_{\langle i,j\rangle}{}\hspace{-0.2em}c_i^{\dagger}c_j+B\hspace{-0.2em} \sum\limits_{i}{}\hspace{-0.2em}c_i^{\dagger}\hat{m}\cdot\vec{s}\hspace{0.2em}c_i\nonumber\\
&&
+it_I\hspace{-0.6em}\sum\limits_{\langle\langle i,j\rangle\rangle}{}\hspace{-0.6em}\nu_{ij}c_i^{\dagger}s_zc_j-it_{IR}\hspace{-0.4em}\sum\limits_{\langle\langle i,j\rangle\rangle}{}\hspace{-0.5em}\mu_{ij}c_i^{\dagger}\left(\vec{s}\times\hat{d}_{i,j}\right)_z\hspace{-0.3em}c_j
\end{eqnarray}
\end{samepage}
\vspace{0.3em}
where $c_i^{\dagger}=\left(c_{i,\uparrow}^{\dagger},c_{i,\downarrow}^{\dagger}\right)$ ; $\nu_{ij} = \frac{(\vec{d_i}\times\vec{d_j})\cdot\hat{z}}{\mid\vec{d_i}\times\vec{d_j}\mid}$ can take values $\pm 1$, $\vec{d}_i$ and $\vec{d}_j$ are the two nearest neighbor bond connecting the next nearest neighbor sites and $\hat{d}_{i,j}$ is the unit vector from $j^{th}$ site to $i^{th}$ site of honeycomb lattice. In buckled honeycomb lattice, $\mu_{ij}$ equals +1 if the lattice point is above a plane and -1 if below.

\begin{figure*}[htbp]
\centering
\includegraphics[scale=1.0]{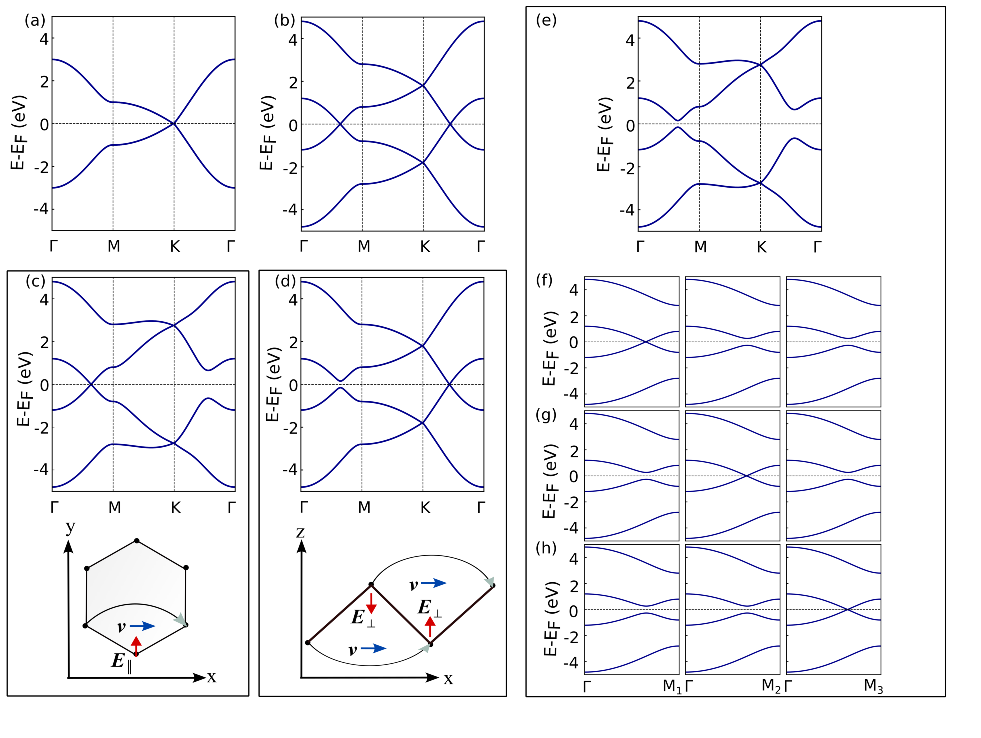}
\caption{Results obtained from the tight binding model. \textbf{(a)} Band structure for $t=1.0$, $B$=$t_{I}$=$t_{IR}=0$. In this case each bands are doubly degenerate and as expected for honeycomb structure, the four-fold degenerate Dirac point is situated at high-symmetry point K. \textbf{(b)} Band structure for $t=1.0$, $B=1.8t$,$\phi=0$, $t_{I}$=$t_{IR}=0$. Here the band degeneracy is removed due to broken TRS. \textbf{(c)} Band structure for $t=1.0$, $B=1.8t$, $\phi=0$, $t_{I}$=$0.4t$, $t_{IR}=0$. A semiclassical explanation of the intrinsic-SOC term in planar honeycomb structure is shown at bottom. \textbf{(d)} Band structure for $t=1.0$, $B=1.8t$, $\phi=0$, $t_{I}=0$, $t_{IR}=0.1t$. A semiclassical explanation of the intrinsic-Rashba SOC term for buckled honeycomb structure is shown at bottom. \textbf{(e)} Band structure for $t=1.0$, $B=1.8t$, $\phi=0$, $t_{I}=0.4t$, $t_{IR}=0.1t$. Band structure along different high symmetry lines are shown for \textbf{(f)} $\phi=30^{\circ},210^{\circ}$, \textbf{(g)} $\phi=90^{\circ},270^{\circ}$ and \textbf{(h)} $\phi=150^{\circ},330^{\circ}$ for $t=1.0$, $B=1.8t$, $t_{I}=0.4t$ and $t_{IR}=0.1t$. The high-symmetry points are same as shown in figure \ref{fig:topology_02}.}
\label{fig:TB}
\end{figure*}

In the following we shall describe the different terms of the Hamiltonian (equation \ref{TB}) and the resulting band dispersion of the TB model. Our results are summarized in figure \ref{fig:TB}. The first term represents the nearest neighbour electron-hopping in a planar honeycomb structure which brings the Dirac points at the corner points $K$ and $K'$ of the first BZ and the electronic band-structure consists of two doubly degenerate bands as presented in figure \ref{fig:TB}(a). 

The effect of the in-plane variation of moment is captured by the second term of the Hamiltonian where $\hat{m} = \left(cos\phi ,sin\phi\right)$ and $\phi$ is the azimuthal angle as introduced above (see figure \ref{fig:topology}(a)). As a consequence of broken TRS, a finite magnetic field $B$ removes the band degeneracy, causing vertical shifts in the band structure for spin-up and spin-down bands as shown in figure \ref{fig:TB}(b). At the Fermi energy $E_F$, there are two two-fold degenerate points, akin to 2D Weyl points (see figure \ref{fig:TB}(b)). Hence, there are total twelve such points in the full BZ. These topological objects differ from conventional 3D Weyl points, adding to their uniqueness \cite{2DWP1,2DWP2}. The separation between these Weyl points at $E_F$ is adjustable through tuning the magnitude of B. For a qualitative comparison with ML OsCl$_3$ band structure, we select $B=1.8t$, positioning one degenerate point between the $\Gamma-M$ line and the other between the $K-\Gamma$ high-symmetry line. Additionally, at the high-symmetry point $K$, inherent band degeneracies persist, which is evident even in presence of SOC, as demonstrated in the DFT band structures and as a consequence, Chern number C shows peak at these degeneracies (for example see figure \ref{fig:topology}(b)).

Without SOC, the TB Hamiltonian breaks TRS ($\hat{T}$) as previously discussed. It also disrupts mirror symmetry along the $\hat{M}_z$ plane due to the in-plane orientation of magnetic moments, yet maintains inversion symmetry ($\hat{I}$). A non-zero Chern number requires breaking both $\hat{T}\otimes\hat{M}_z$ and $\hat{T}\otimes\hat{M}_z\otimes\hat{I}$ symmetries. The presence of $\hat{T}\otimes\hat{M}_z$ symmetry renders Berry curvature an odd function of k within the first BZ, leading to a null Chern number upon integration. With $\hat{T}\otimes\hat{M}_z\otimes\hat{I}$ symmetry, Berry curvature vanishes at each k-point \cite{IPQAHE3}. Consequently, the TB model, without the last two terms in equation \ref{TB}, is topologically trivial.

The third and fourth terms of the Hamiltonian, arising from SOC, play a crucial role in forming an insulating non-trivial topological states. A semiclassical perspective aids in understanding these terms, offering insights into the behavior of the system and guiding material design.

In this approach, heavy transition metal ions at honeycomb lattice-points serve as electric field sources for electrons. This field divides into two components: $E_{||}$, parallel to the honeycomb plane, and $E_{\perp}$, perpendicular to it. Without buckling, the $E_{\perp}$ is expected to be trivially zero. However, buckling or in-plane mirror symmetry breaking assisted by ligands, generating a nonzero $E_{\perp}$, as illustrated in the inset figures of \ref{fig:TB}(c) and (d). Notice that, although in-plane orientation of moments also breaks the $\hat{M}_z$ mirror symmetry, it can not provide nonzero $E_{\perp}$.

Electron-hopping between sites, as influenced by this electric field, generates an effective magnetic field $\vec{B}_{eff}\propto\vec{v}\times\vec{E}$, where $\vec{v}$ is the electron velocity. This field interacts with electron spins according to the effective Hamiltonian $\hat{H}_{eff}=-\vec{s}\cdot\vec{B}_{eff}$. In the honeycomb structure, SOC vanishes for nearest-neighbour hopping due to the honeycomb geometry but significant for next-nearest-neighbour hopping. $E_{||}$ introduces the third term, known as intrinsic-SOC term (illustrated in the inset of figure \ref{fig:TB}(c)), while $E_{\perp}$, unlike the Rashba effect where an electric field is applied externally to 2D electrons, arises intrinsically either in the buckled structure or in a tilted octahedral environment as provided by the Cl-ligands in ML OsCl$_3$ (shown in the inset of figure \ref{fig:TB}(d)). In-plane mirror symmetry breaking is crucial for a nonzero $E_{\perp}$ and, thus enabling the intrinsic-Rashba effect, as included in the fourth term of the TB model (equation \ref{TB}).

The direction of electron hopping alters the orientation of $\vec{B}_{eff}$, impacting the ground state of $\hat{H}_{eff}$, which depends on both $\vec{B}_{eff}$ and spin directions. For intrinsic-SOC, this results in the $\nu_{ij}$ term. Additionally, for intrinsic-Rashba SOC (figure \ref{fig:TB}(d) inset) where the next-nearest neighbour electron hopping experiences $E_{\perp}$ differently above and below the plane, leading to the $\mu_{ij}$ term in the fourth term of equation \ref{TB}.

Figure \ref{fig:TB}(c) illustrates the impact of intrinsic-SOC without the $t_{IR}$ term in equation \ref{TB}. The third term maintains all the key symmetries ($\hat{T}$, $\hat{M}_z$ and $\hat{I}$), hence resulting in a Chern number C=0. Here, the degeneracy at the Fermi energy $E_F$ along the $K-\Gamma$ line is lifted, with the gap size dependent on the strength of $t_{I}$, while the degeneracy along the $\Gamma-M$ line remains. Conversely, incorporating intrinsic-Rashba SOC with $t_{I}=0$ in equation \ref{TB} leads to the opposite effect: it opens a gap along the $\Gamma-M$ path, removing degeneracy there, but the degeneracy along $K-\Gamma$ line persists as depicted in figure \ref{fig:TB}(d). However, intrinsic-Rashba SOC breaks $\hat{M}_z$ symmetry while preserving $\hat{T}$ and $\hat{I}$, introducing non-trivial topology to the system. 

The TB band-structure, including all terms, is shown in figure \ref{fig:TB}(e), exhibiting insulating behavior with $C=-1$ for $\phi=0$. This four-band generic model aligns with the $J_{eff}=1/2$ manifold near $E_F$ for ML OsCl$_3$. The presence of mirror plane $\hat{M}_i$ in a system nullifies the corresponding non-diagonal conductivity tensor components $\sigma_{ij}$ \cite{IPQAHE2}. In ML OsCl$_3$, both the in-plane orientation of moments and tilted octahedra formed by Cl-ligands break the in-plane mirror symmetry. The out-of-plane mirror symmetry is broken for particular in-plane orientations of magnetic moments. However, when moments are oriented perpendicular to the out-of-plane mirrors, the mirror symmetry is retained in the system and as a consequence the $\sigma_{xy}$ should vanish. As discussed above, the presence of mirror symmetry in the ML OsCl$_3$ maintains the degeneracy at $E_F$ along the $\Gamma-M$ path, marking the non-adiabatic phases where topological phase transitions occur. For the TB model also, the out-of-plane mirror symmetry is preserved at $\phi$ = $30^{\circ}$, $90^{\circ}$, $150^{\circ}$, $210^{\circ}$, $270^{\circ}$ and $330^{\circ}$, paralleling findings with the DFT calculation. Thus, the topological phase transition points and band-degeneracy properties due to MAE in the TB model (figure \ref{fig:TB}(f)-(h)) and DFT results for ML OsCl$_3$ (figure \ref{fig:topology_02}) are consistent.

\section{Conclusion}
\label{IV}

We have demonstrated the change in electronic topology of ML OsCl$_3$ with variation in in-plane moments where the tilted Cl-ligands provide an effective buckling effect and depending upon the moment orientation, the out-of-plane mirror symmetry can be preserved or broken, leading to the change in topological phase as shown in figure \ref{fig:topology}(a). The symmetry driven easy-plane magnetic anisotropy helps to change the topological phases where the broken $\hat{M}_z$ mirror symmetry is incorporated in intrinsic Rashba-SOC with in-plane orientation of moments. As a consequence the C changes from -1 to +1 several times by tuning the moment orientation. This highlights a significant interplay between electronic band topology and magnetism. It is important to note that these topological properties can be captured by a generic TB model as it does not depend on the $J_{eff}=1/2$ orbital physics of ML OsCl$_3$. However, increasing the Hubbard-U value brings orbital physics into play, potentially transforming the system from an in-plane ferromagnet to a Kitaev spin liquid state \cite{QSL} and this transition may disrupt the topology of the electronic band structure \cite{Oc}. Similar phenomena might be observed in ML RuX$_3$ (where X = Cl, Br or I) due to their strong SOC and presence of relevant symmetries. Our study sheds light on the tunable electronic topology of planar honeycomb structures with in-plane ferromagnetism and points to the importance of the role of ligands in achieving non-trivial topology. These findings could be crucial for developing materials with specific electronic properties.

\ack
R.D. would like to thank the Council of Scientific and Industrial Research (CSIR), India for research fellowship (File No. 09/080(1171)/2020-EMR-I). I.D. thanks Science
and Engineering Research Board (SERB) India (Project No. CRG/2021/003024) and Technical Research Center, Department of Science and Technology Government of India for support.
\section*{References}
\providecommand{\newblock}{}


\begin{thebibliography}{10}
\expandafter\ifx\csname url\endcsname\relax
  \def\url#1{{\tt #1}}\fi
\expandafter\ifx\csname urlprefix\endcsname\relax\def\urlprefix{URL }\fi
\providecommand{\eprint}[2][]{\url{#2}}

\bibitem{QHE1}
Klitzing K~v, Dorda G and Pepper M 1980 {\em Phys. Rev. Lett.\/} {\bf 45}(6)
  494--497 \urlprefix\url{https://link.aps.org/doi/10.1103/PhysRevLett.45.494}

\bibitem{QHE2}
Laughlin R~B 1981 {\em Phys. Rev. B\/} {\bf 23}(10) 5632--5633
  \urlprefix\url{https://link.aps.org/doi/10.1103/PhysRevB.23.5632}

\bibitem{QHE3}
Thouless D~J, Kohmoto M, Nightingale M~P and den Nijs M 1982 {\em Phys. Rev.
  Lett.\/} {\bf 49}(6) 405--408
  \urlprefix\url{https://link.aps.org/doi/10.1103/PhysRevLett.49.405}

\bibitem{RMP_Berry}
Xiao D, Chang M~C and Niu Q 2010 {\em Rev. Mod. Phys.\/} {\bf 82}(3) 1959--2007
  \urlprefix\url{https://link.aps.org/doi/10.1103/RevModPhys.82.1959}

\bibitem{RMP_AHE1}
Nagaosa N, Sinova J, Onoda S, MacDonald A~H and Ong N~P 2010 {\em Rev. Mod.
  Phys.\/} {\bf 82}(2) 1539--1592
  \urlprefix\url{https://link.aps.org/doi/10.1103/RevModPhys.82.1539}

\bibitem{RMP_AHE2}
Chang C~Z, Liu C~X and MacDonald A~H 2023 {\em Rev. Mod. Phys.\/} {\bf 95}(1)
  011002 \urlprefix\url{https://link.aps.org/doi/10.1103/RevModPhys.95.011002}

\bibitem{Haldane}
Haldane F~D~M 1988 {\em Phys. Rev. Lett.\/} {\bf 61}(18) 2015--2018
  \urlprefix\url{https://link.aps.org/doi/10.1103/PhysRevLett.61.2015}

\bibitem{BBC1}
Halperin B~I 1982 {\em Phys. Rev. B\/} {\bf 25}(4) 2185--2190
  \urlprefix\url{https://link.aps.org/doi/10.1103/PhysRevB.25.2185}

\bibitem{BBC2}
Dugaev V~K, Barnaś J, Taillefumier M, Canals B, Lacroix C and Bruno P 2008
  {\em Journal of Physics: Conference Series\/} {\bf 104} 012018
  \urlprefix\url{https://dx.doi.org/10.1088/1742-6596/104/1/012018}

\bibitem{Application1}
Hasan M~Z and Kane C~L 2010 {\em Rev. Mod. Phys.\/} {\bf 82}(4) 3045--3067
  \urlprefix\url{https://link.aps.org/doi/10.1103/RevModPhys.82.3045}

\bibitem{Application2}
He Q~L, Hughes T~L, Armitage N~P, Tokura Y and Wang K~L 2022 {\em Nature
  Materials\/} {\bf 21}(1) 15--23
  \urlprefix\url{https://doi.org/10.1038/s41563-021-01138-5}

\bibitem{Application3}
Deng Y 2022 {\em Journal of Physics: Conference Series\/} {\bf 2386} 012058
  \urlprefix\url{https://doi.org/10.1088/1742-6596/2386/1/012058}

\bibitem{OPQAHE1}
Yu R, Zhang W, Zhang H~J, Zhang S~C, Dai X and Fang Z 2010 {\em Science\/} {\bf
  329} 61--64 (\textit{Preprint}
  \eprint{https://www.science.org/doi/pdf/10.1126/science.1187485})
  \urlprefix\url{https://www.science.org/doi/abs/10.1126/science.1187485}

\bibitem{OPQAHE2}
Wang J, Lian B, Zhang H, Xu Y and Zhang S~C 2013 {\em Phys. Rev. Lett.\/} {\bf
  111}(13) 136801
  \urlprefix\url{https://link.aps.org/doi/10.1103/PhysRevLett.111.136801}

\bibitem{OPQAHE3}
Dolui K, Ray S and Das T 2015 {\em Phys. Rev. B\/} {\bf 92}(20) 205133
  \urlprefix\url{https://link.aps.org/doi/10.1103/PhysRevB.92.205133}

\bibitem{OPQAHE4}
You J~Y, Zhang Z, Gu B and Su G 2019 {\em Phys. Rev. Appl.\/} {\bf 12}(2)
  024063
  \urlprefix\url{https://link.aps.org/doi/10.1103/PhysRevApplied.12.024063}

\bibitem{AHAFM}
Šmejkal L, MacDonald A~H, Sinova J, Nakatsuji S and Jungwirth T 2022 {\em
  Nature Reviews Materials\/} {\bf 7} 482--496 ISSN 2058-8437
  \urlprefix\url{https://doi.org/10.1038/s41578-022-00430-3}

\bibitem{Landau}
Landau L~D and Lifshitz E~M 1984 {\em Electrodynamics of Continuous Media\/}
  2nd ed Course of Theoretical Physics (Butterworth-Heinemann) ISBN
  978-0-7506-2634-7
  \urlprefix\url{https://www.elsevier.com/books/electrodynamics-of-continuous-media/landau/978-0-7506-2634-7}

\bibitem{IPQAHE1}
Zhang Y and Zhang C 2011 {\em Phys. Rev. B\/} {\bf 84}(8) 085123
  \urlprefix\url{https://link.aps.org/doi/10.1103/PhysRevB.84.085123}

\bibitem{IPQAHE2}
Liu X, Hsu H~C and Liu C~X 2013 {\em Phys. Rev. Lett.\/} {\bf 111}(8) 086802
  \urlprefix\url{https://link.aps.org/doi/10.1103/PhysRevLett.111.086802}

\bibitem{IPQAHE3}
Ren Y, Zeng J, Deng X, Yang F, Pan H and Qiao Z 2016 {\em Phys. Rev. B\/} {\bf
  94}(8) 085411
  \urlprefix\url{https://link.aps.org/doi/10.1103/PhysRevB.94.085411}

\bibitem{IPQAHE4}
Liu Z, Zhao G, Liu B, Wang Z~F, Yang J and Liu F 2018 {\em Phys. Rev. Lett.\/}
  {\bf 121}(24) 246401
  \urlprefix\url{https://link.aps.org/doi/10.1103/PhysRevLett.121.246401}

\bibitem{IPQAHE5}
Guo X, Liu Z, Liu B, Li Q and Wang Z 2020 {\em Nano Letters\/} {\bf 20}
  7606--7612 pMID: 32852221 (\textit{Preprint}
  \eprint{https://doi.org/10.1021/acs.nanolett.0c03136})
  \urlprefix\url{https://doi.org/10.1021/acs.nanolett.0c03136}

\bibitem{Oc}
Sheng X~L and Nikoli\ifmmode~\acute{c}\else \'{c}\fi{} B~K 2017 {\em Phys. Rev.
  B\/} {\bf 95}(20) 201402
  \urlprefix\url{https://link.aps.org/doi/10.1103/PhysRevB.95.201402}

\bibitem{Recent}
Li Z, Han Y and Qiao Z 2022 {\em Phys. Rev. Lett.\/} {\bf 129}(3) 036801
  \urlprefix\url{https://link.aps.org/doi/10.1103/PhysRevLett.129.036801}

\bibitem{PAW}
Bl\"ochl P~E 1994 {\em Phys. Rev. B\/} {\bf 50}(24) 17953--17979
  \urlprefix\url{https://link.aps.org/doi/10.1103/PhysRevB.50.17953}

\bibitem{VASP}
Kresse G and Hafner J 1993 {\em Phys. Rev. B\/} {\bf 47}(1) 558--561
  \urlprefix\url{https://link.aps.org/doi/10.1103/PhysRevB.47.558}

\bibitem{GGA}
Perdew J~P, Burke K and Ernzerhof M 1996 {\em Phys. Rev. Lett.\/} {\bf 77}(18)
  3865--3868
  \urlprefix\url{https://link.aps.org/doi/10.1103/PhysRevLett.77.3865}

\bibitem{W90_1}
Pizzi G, Vitale V, Arita R, Blügel S, Freimuth F, Géranton G, Gibertini M,
  Gresch D, Johnson C, Koretsune T, Ibañez-Azpiroz J, Lee H, Lihm J~M,
  Marchand D, Marrazzo A, Mokrousov Y, Mustafa J~I, Nohara Y, Nomura Y,
  Paulatto L, Poncé S, Ponweiser T, Qiao J, Thöle F, Tsirkin S~S, Wierzbowska
  M, Marzari N, Vanderbilt D, Souza I, Mostofi A~A and Yates J~R 2020 {\em
  Journal of Physics: Condensed Matter\/} {\bf 32} 165902
  \urlprefix\url{https://dx.doi.org/10.1088/1361-648X/ab51ff}

\bibitem{W90_2}
Marzari N, Mostofi A~A, Yates J~R, Souza I and Vanderbilt D 2012 {\em Rev. Mod.
  Phys.\/} {\bf 84}(4) 1419--1475
  \urlprefix\url{https://link.aps.org/doi/10.1103/RevModPhys.84.1419}

\bibitem{WT}
Wu Q, Zhang S, Song H~F, Troyer M and Soluyanov A~A 2018 {\em Computer Physics
  Communications\/} {\bf 224} 405--416 ISSN 0010-4655
  \urlprefix\url{https://www.sciencedirect.com/science/article/pii/S0010465517303442}

\bibitem{IBC}
Tu M~W~Y, Li C, Yu H and Yao W 2020 {\em 2D Materials\/} {\bf 7} 045004
  \urlprefix\url{https://dx.doi.org/10.1088/2053-1583/ab89e8}

\bibitem{WHSM_1}
You J~Y, Chen C, Zhang Z, Sheng X~L, Yang S~A and Su G 2019 {\em Phys. Rev.
  B\/} {\bf 100}(6) 064408
  \urlprefix\url{https://link.aps.org/doi/10.1103/PhysRevB.100.064408}

\bibitem{WHSM_2}
Jiao Y, Zeng X~T, Chen C, Gao Z, Song K, Sheng X~L and Yang S~A 2023 {\em Phys.
  Rev. B\/} {\bf 107}(7) 075436
  \urlprefix\url{https://link.aps.org/doi/10.1103/PhysRevB.107.075436}

\bibitem{HDP}
Yu Y, Xie X, Liu X, Li J, Peeters F~M and Li L 2022 {\em Applied Physics
  Letters\/} {\bf 121} 112405 \urlprefix\url{https://doi.org/10.1063/5.0105605}

\bibitem{Main_tb}
Liu C~C, Jiang H and Yao Y 2011 {\em Phys. Rev. B\/} {\bf 84}(19) 195430
  \urlprefix\url{https://link.aps.org/doi/10.1103/PhysRevB.84.195430}

\bibitem{2DWP1}
Yoshida H, Zhang T and Murakami S 2023 {\em Phys. Rev. B\/} {\bf 107}(3) 035122
  \urlprefix\url{https://link.aps.org/doi/10.1103/PhysRevB.107.035122}

\bibitem{2DWP2}
Yoshida H, Zhang T and Murakami S 2023 {\em Phys. Rev. B\/} {\bf 108}(7) 075160
  \urlprefix\url{https://link.aps.org/doi/10.1103/PhysRevB.108.075160}

\bibitem{QSL}
Gu Q, Pandey S~K and Lin Y 2023 {\em arXiv:2304.04257 [cond-mat]\/}
  \urlprefix\url{https://arxiv.org/abs/2304.04257}

\end{thebibliography}
\end{document}